\documentclass[aps,prd,12pt]{revtex4-1}
\usepackage{amsmath,graphicx}
\usepackage{bm}
\usepackage{amssymb}
\usepackage{graphicx,xcolor}
\usepackage{epigraph}
\usepackage{csquotes}
\usepackage{amsmath,amssymb,stmaryrd}
\begin{document}
\def\e{\enquote}
\def\tr{\rm{Tr}}
\def\la{{\langle}}
\def\ra{{\rangle}}
\def\a{{\alpha}}
\def\q{\quad}
\def\w{\tilde}
\def\om{\omega_L}
\def\t{\tilde{t}}
\def\a{\hat{A}}
\def\h{\hat{H}}
\def\E{\mathcal{E}}\def\la{{\langle}}
\def\u{\hat U}
\def\U{\hat U}
\def\C{\hat C}
\def\D{\hat D}
\def\S{\tilde S}
\def\A{{\textbf A}}
\def\AA{{\tilde A}}
\def\B{{\textbf B}}
\def\Delt{\tilde \Delta}
\def\QQ{\hat S}
\def\R{\text {Re}}
\def\I{\text {Im}}
\def\e{\enquote}
\def\qq{s}
\def\Q{S}
\def\fb{\overline F}
\def\wb{\overline W}
\def\nl{\newline}
\def\h{\hat H}
\def\ff{\overline q}
\def\k{\overline k}
\def\F {Q}
\def\f{q}
\def\lm{\lambda}
\def\lmu{\underline\lambda}
\def\q{\quad}
\def\t{\tau}
\def\l{\ell}
\def\n{\\ \nonumber}
\def\ra{{\rangle}}
\def\Ep{{\mathcal{E}}}
\def\T{T_{total}}
\def\M{{\mathcal{M}}}
\def\omga{{\epsilon}}
\def\t{{\tau_{SW}}}
\def\h{\hat{H}}
\title{Tunnelling times,  Larmor clock, and the elephant in the room}
%
%

\author {D. Sokolovski$^{1,3}$} 
\email {dgsokol15@gmail.com}
\author {E. Akhmatskaya$^{2,3}$ } 
\affiliation{$^1$ Departmento de Qu\'imica-F\'isica, Universidad del Pa\' is Vasco, UPV/EHU, Leioa, Spain}
\affiliation {$^2$ Basque Center for Applied Mathematics (BCAM),\\ Alameda de Mazarredo 14, 48009 Bilbao, Bizkaia, Spain}
\affiliation{$^3$ IKERBASQUE, Basque Foundation for Science, Plaza Euskadi 5, 48009 Bilbao, Spain}\date{\today}

\begin{abstract}
\begin{center}
{\bf ABSTRACT:}
\end{center}
\noindent
{A controversy surrounding the \e{tunnelling time problem} stems from the seeming inability of quantum mechanics to provide, in the usual way, a definition of the duration a particle is supposed to spend in a given region of space. For this reason the problem is often approached from an \e{operational} angle. Typically, one tries to mimic, in a quantum case, an experiment which yields the desired result for a classical particle. One such approach is based on the use of a Larmor clock. We show that the difficulty with applying a non-perturbing Larmor clock in order to \e{time} a classically forbidden transition arises from the quantum Uncertainty Principle. We also demonstrate that for this reason a Larmor time (in fact, none of the Larmor times) cannot be interpreted as a physical time interval. We also provide a theoretical description of the quantities measured by the clock.}  
\end{abstract}

\maketitle

\section{Introduction}
The \e{tunnelling time} problem which has been with us for nearly a century \cite{McColl}, 
still has its share of controversy (for a recent review see \cite{Rev}), and for a good reason. A prerequisite for any constructive discussion is a possibility to define its subject in a meaningful way.
For a classical particle, a duration spent in a given region of space is indeed a well established and useful concept.
In quantum mechanics, the Uncertainty Principle (UP) forbids answering the "which way?" question if two or more pathways leading to the same final outcome interfere \cite{FeynL}. By the same token a duration, readily determined for each path, must remain indeterminate for a process 
where interference plays a crucial role. This is particularly true in the case of tunnelling. 
\newline
The early attempts to define the duration a quantum particle spends in the barrier by following the evolution of the transmitted wave packet
\cite{WE}-\cite{Smith} yielded the so-called Wigner-Smith (WS) time delay, essentially the energy derivative of the phase of the transmission amplitude. One immediate problem with the method is that if the WS result is used to estimate the time spent by the particle in the barrier, 
this time turns out to be shorter than the barrier width divided by the speed of light. This apparently  \e{superluminal behaviour}
does not lead to a conflict with Einstein's relativity for the simple reason that, in accordance with the Uncertainty Principle, the WS time cannot be interpreted as a physical time interval spent by a tunnelling particle in the barrier \cite{DSnat}. 
However, as was noted in \cite{Rev}, the argument of \cite{DSnat} applies to the \e{phase time} of \cite{WE}-\cite{Smith}.
Would it still be true if the tunnelling time were defined in a different manner?
\newline  
An alternative approach was proposed by A.I. Baz' \cite{Baz1}, 
who employed Larmor precession of a magnetic moment (spin) in a magnetic field, small enough not to affect tunnelling seriously \cite{Baz2}.
The interest in the Larmor (Baz') clock was recently renewed after its experimental realisation was reported in \cite{Stein}, and in what follows we will analyse it in some detail. By construction, such a clock probes the response of a scattering amplitude
to a small variation of the potential, rather than to a variation of the particle's energy. 
Thus, the Larmor time was found to disagree with the Wigner and Smith result, and proposed to be the \e{correct} estimate of the duration of a scattering process (see the footnote on p.169 of \cite{Baz3}).
Despite Baz's  assertion in \cite{Baz3},  the Larmor clock approach soon encountered its own difficulties. 
In particular, if applied to tunnelling transmission the method yielded not one but two time parameters, which B\"uttiker \cite{Butt} proposed to combine into 
a single \e{interaction time.} In \cite{SB} Sokolovski and Baskin have shown the two Larmor times to be the real and imaginary parts of 
a \e{complex time} obtained as an average, in which the usual probabilities were replaced with quantum probability amplitudes. 
The lack of clarity about these matters points to a more fundamental problem, which requires further attention.
\newline
The purpose of this paper is to demonstrate that the difficulty in deducing the duration spent in the barrier,  evident in the analysis of the Wigner-Smith time delay \cite{DSnat}, persists also in the conceptually different  Larmor clock approach \cite{Baz1}-\cite{SB}. 
To do so we will again appeal to the Uncertainty Principle, a rule of primary importance for any discussion of the tunnelling 
time problem, yet rarely mentioned in such discussions. It will also be upon us to answer the question \e{does a Larmor clock measure a physical time interval and, if not, then what does it measure?}

\section{Results}
To lay bare the conceptual difficulty, we start by considering a simple thought experiment, where an electron, with its spin polarised along the $x$-axis, enters an interferometer shown in Fig.1
in a wave packet state $|G_0\ra$, and is detected after exiting the second beam splitter, as shown in Fig.1. Travelling via different arms of the interferometer, the electron spends different durations, $\tau_1$ and $\tau_2$, in a region containing constant magnetic field directed along the $z$-axis, $B$  (in an experiment using photons and 
Faraday's rotation the field would be directed along the arms). An additional element (e.g., an extra potential) in the second arm 
ensures that an extra phase, $\phi$ is acquired there by both spin components.  So how much time did the electron spend in the magnetic field?
\begin{figure}[h]
\includegraphics[angle=0,width=12cm, height= 8cm]{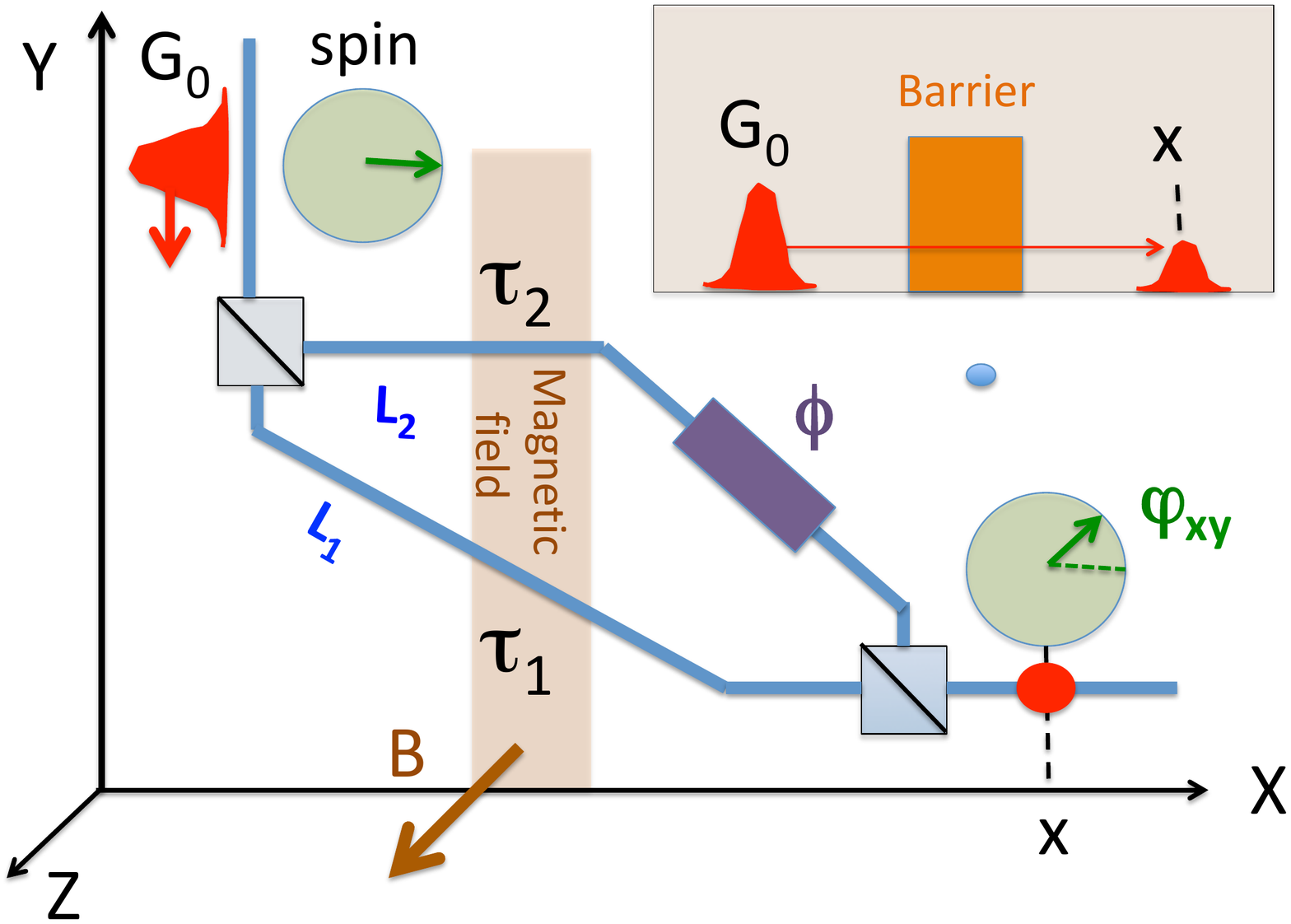}
\caption {A particle reaches the final position $x$ after passing through an interferometer, 
and a weekly coupled Larmor clock is used to determine the duration it spends in the magnetic field. 
The case of tunnelling across a potential barrier, shown in the inset, is more complicated, yet conceptually similar.}
\end{figure}
\newline
The question is more difficult that it may seem. If the wave packet travelling at a velocity $v$ is fast, and the field is not too strong, the two spin components acquire, in each arm,  
phases $\exp(\pm \om \tau_{1,2})$, where $\om$ is the  Larmor frequency.
Thus, beyond the second beam splitter the wave function is given by (the $\hat \sigma$s are Pauli matrices)
\begin{eqnarray}\label{1}
\la x|\Phi\ra=[G_1(x,t)\exp(-i\om \hat \sigma_z \tau_1) +\n
G_2(x,t)\exp(-i\om \hat \sigma_z \tau_2)]|\uparrow_x\ra,
\end{eqnarray}
where $G_{1,2}(x,t)$ are the parts of the original wave packets arriving at $x$ via the first and the second arm, respectively. 
One notes that the sum of the rotations in the square brackets does not add up to a single rotation around the $z$-axis, so 
no duration can be deduced from Eq.(\ref{1}) directly. Perhaps, making the field small could help? Indeed, sending $\om \to 0$ 
and keeping only the linear terms, one finds $\la x|\Phi\ra\approx [G_1(x,t)+G_2(x,t)] (1 -i\om \sigma_z \overline{\tau})|\uparrow_x\ra$,
which now looks like an overall rotation through a small  angle $ \om \overline{\tau}$.
Does this mean that
\begin{eqnarray}\label{2}
\overline{\tau}\equiv \frac{\tau_1G_1(x,t)+\tau_2G_2(x,t)}{G_1(x,t)+G_2(x,t)}\equiv \tau_1\alpha(\tau_1)+\tau_2\alpha(\tau_2)
\end{eqnarray}
is  a suitable candidate for the duration spent in the field? Not quite so. The quantities $G_{i}$ are the transition amplitudes \cite{FeynL} 
for an electron, initially in $|G_0\ra$, to reach $|x\ra$ via the $i$-th arm of the interferometer, and $\overline{\tau}$ is complex valued.
This new problem can be dealt with by evaluating the mean angle of precession in the $xy$-plane, $\varphi_{xy}$, guaranteed at least to be real.
The result,
\begin{eqnarray}\label{3}
\varphi_{xy}\approx \la \Phi|x\ra \hat \sigma_y\la x|\Phi\ra=\om\text{Re}[\overline{\tau}], 
\end{eqnarray}
appears to give preference to the real part of $\overline{\tau}$, and may look satisfactory. 
(Note that measuring the angle of rotation in the $xz$-plane would yield also the value of $\text{Im}[\overline{\tau}]$, but it is not important to us here.)
\newline 
However, our real problems are only beginning. A {\it non-negative} probability distribution, $\rho(z) \ge 0$, has many useful properties. For example, 
an expectation value $\la z\ra$  marks roughly the centre of the region where $\rho(z)\ne 0$, and the variance gives an estimate of  the size of this region.
This is no longer true for the distributions which change sign, and the \e{average} in Eq.(\ref{2}) is of this latter type. 
Adjusting the phases and lengths, one can ensure that $G_2(x,t)\approx -G_1(x,t)$, and make the denominator in Eq.(\ref{2}) small.
A similar cancellation will not occur in the numerator, and $\overline{\tau}$ can be made as large as one wants. 
On the other hand, with both arms of about the same length, $\L_1\approx L_2\approx L$, the electron spends in motion approximately $\sim L/v$. 
Now the \e{duration} in Eq.(\ref{3}) can easily exceed the total time electron was in motion, 
\begin{eqnarray}\label{4}
\text{Re}[\overline{\tau}] >>L/v\equiv T_{total}. 
\end{eqnarray}
Similarly, $\tau_{1,2}$ and $G_{1,2}$ could be chosen so that $\overline{\tau}=0$, making it look like the electron, known
to move at a speed $v$ in each arm, crosses the field infinitely fast if both arms are considered together. 
These are serious issues, which should not be ignored. One has to decide whether to allow a quantum particle to spend more time that it has at its disposal, and hail Eq.(\ref{4}) as a new triumph of quantum theory. The other possibility is to conclude that something is  wrong with the very question asked.
It is, indeed, frustrating to have two durations, $\tau_1$ and $\tau_2$, and to be unable to combine them 
into anything meaningful if a particle passes through both arms of  the interferometer in Fig.1. 
\newline
The frustration is of a familiar kind. In a Young's double-slit experiment, an electron passes trough one of the two slits, but it is not possible to know which particular slit was chosen. The impossibility of answering the \e{which way?} question, without destroying interference,  is the essence of the Uncertainty Principle, without which 
quantum mechanics {\it \e{would collapse}} \cite{FeynL}. The experiment in Fig.1 is a kind of a double-slit case, with the only difference
that the \e{which way?} question has been disguised as a \e{how much time?} query. 
\newline
It is instructive to see how quantum mechanics implements
the Principle in practice. Since the only {\it a priory} restriction on the in general complex valued {\it relative amplitudes} $\alpha_{1,2}$ in Eq.(\ref{2}) is that they should add up to unity, $\alpha_1+\alpha_2=1$, one can find suitable $\alpha$s for {\it any} choice of a complex  $\overline{\tau}$, 
\begin{eqnarray}\label{4a}
\alpha_1=\frac{\overline{\tau}-\tau_2}{\tau_1-\tau_2}, \q \alpha_2=-\frac{\overline{\tau}-\tau_1}{\tau_1-\tau_2}.
\end{eqnarray}
Unable to forbid one to ask the question operationally, quantum theory gives all possible answers, suitable and unsuitable, 
according to the circumstances. Depending on the parameters of the interferometer, the measured real part of $\overline{\tau}$ can be positive, negative, zero, coincide with $\tau_1$ or $\tau_2$, or lie between them.
The answer to a question that should not have an answer can be {\it \e{anything at all}}.
\newline
One can envisage a following dialogue between an experimentalist Alice ($\A$) and a theoretician Bob ($\B$).

$\A$: I have just measured the mean angle $\varphi_{xy}$, and divided it by the Larmor frequency. It follows that $\tau_{Alice}=\text{Re}[\overline{\tau}]= (\tau_1+\tau_2)/2$, a perfectly reasonable result. 
And I was told this time does not exist. 
\newline
 $\B$: It does not. Change your settings, and the same procedure will give you $\text{Re}[\overline{\tau}] <0$.
The time parameter you measure is not a meaningful duration.

$\A$:  Let us just forget about the cases where something goes wrong. Surely, in my case it {\it is} the time an electron spends in the magnetic field. 
\newline $\B$: Just don't tell that Carol-the-engineer. What she wants, is a time scale for changing the setup slowly enough for the electron 
\e{to see} its conditions \e{frozen} during its  journey to the detector. For your $\tau_{Alice}$ to serve as a classical time scale you would also need 
to show that $\tau_1^n \alpha(\tau_1)+\tau_2^n \alpha(\tau_2)=\overline{\tau}^n$, $n=1,2,...$. However, this happens only if one of the $\alpha$s
vanishes, in which case either $\tau_1$ or $\tau_2$ is the time scale Carol would be happy with.

$\A$: But this time scale is a very well known and useful concept. How can it not exist?
\newline
 $\B$: It is also an essentially classical concept, useful when there is no interference involved. Make one arm of the interferometer much longer than the other, so that the two parts of the wave packet do not overlap at $x$. Then, at a given $t$, you will know which way the electron has travelled, and also the duration, $\tau_1$ {\it or} $\tau_2$ it has spent in the magnetic field. 
 But then, of course, it would be a {\it different} experiment. 

$\A$: And what if I take instead the imaginary part, or the modulus of $\overline{\tau}$, as was suggested, for example by B\"uttiker \cite{Butt}?
\newline
$\B$: Or any real valued combination of $\text{Re}[\overline{\tau}]$ and $\text{Im}[\overline{\tau}]$. You will still encounter 
\e{times} which are too long for common sense or too short for Einstein's relativity, although with $\tau_{Alice} =|\overline{\tau}|$ you would not need to worry about negative durations. 

$\A$: So what is my \e{time} good for?
\newline
$\B$: It does describe the response of the electron to a small perturbation of a {\it particular type}, a small rectangular potential, 
introduced by the constant magnetic field.  A different \e{time} would arise
if the response to a small oscillating potential were to be studied instead \cite{BL}.

$\A$: So, if my time is not a \e{meaningful duration}, what is it? It looks like one of the \e{weak values} we heard so much about recently \cite{WVrev}.
\newline
$\B$: It is just what Eq.(\ref{3}) says, $\overline{\tau}$ is a sum of relative probability amplitudes for reaching the detector via different arms, 
multiplied by the corresponding durations spent in the field, thus, also an amplitude. And so is every other \e{weak value} \cite{DSann}. Your time is just the real part of a particular probability amplitude. 

$\A$: But I have just measured it.
\newline
$\B$: Not quite, you just measured the spin, and then tried to learn something about electron's translational degree of freedom.
In doing so, you relied on the first-order perturbation theory. Response of a system to a small perturbation is commonly 
described in terms of real valued combinations of the system's probability amplitudes.

$\A$: And what is then an amplitude?
\newline
$\B$: According to Feynman \cite{FeynL}, it is a basic concept in our description of quantum behaviour. 

$\A$: This does not tell me very much. Can you be more specific?
\newline
$\B$: I am afraid not. Nor, I suspect, can anybody else, unless a radically new insight into physics of the double-slit experiment
is gained in future. In Feynman's words, at the moment {\it \e{no one will give you any deeper description of the situation}} \cite{FeynL}.
%

The case of Ref.\cite{Stein} is similar to the one just discussed, if not more involved (see Methods). In Fig.1, there are only two routes by which an electron, starting in a state $|G_0\ra$, can reach 
the final position $x$, and the corresponding amplitude has two components,
\begin{eqnarray}\label{5}
A(x\gets G_0)=G_1(x,t)+G_2(x,t)\equiv \n
A(x\gets G_0|\tau_1)+A(x\gets G_0|\tau_2). 
\end{eqnarray}
For a quantum particle crossing a potential barrier, there are many 
possible $\tau$s,
and many components to the transition amplitude \cite{PRA},
\begin{eqnarray}\label{6}
A(x\gets G_0)=\int_0^{T_{total}} A(x\gets G_0|\tau) d\tau.
\end{eqnarray}
The mean angle of spin's rotation in a small magnetic field, confined to the barrier, is given by an analogue of (\ref{3}) 
\begin{eqnarray}\label{7}
\varphi_{xy}\approx \om\text{Re}\left [\frac{\int_0^{T_{total}} \tau A(x\gets G_0|\tau) d\tau}{\int_0^{T_{total}} A(x\gets G_0|\tau) d\tau} \right]\q\n
=  \om\text{Re}\left [{\int_0^{T_{total}}  \tau \alpha(\tau) d\tau}\right]\equiv  
\om\text{Re}\left [\overline \tau\right]. 
\end{eqnarray}
In the classical limit, highly oscillatory $A(x\gets G_0|\tau)$ develops a stationary region around the classical duration 
$\tau_{class}$, where it varies more slowly. This is the only region contributing to the integral in (\ref{7}), and one recovers the classical result,  
$\overline \tau=\tau_{class}$. But this well defined duration disappears already if $A(x\gets G_0|\tau)$ has two, rather than just one, stationary regions, 
and we are back to the situation similar to the one shown in Fig.1.
\newline
Quantum {\it tunnelling} is a destructive interference phenomenon, where $A(x\gets G_0|\tau)$ in Eq.(\ref{5}) has no stationary regions, and rapidly oscillates
throughout the allowed range $0 \le \tau \le T_{total}$. 
 The tunnelling amplitude (\ref{5}) is extremely  small for a tall or a wide barrier
(see the inset in Fig.1). This happens not because $A(x\gets G_0|\tau)$ is itself small, 
but because its oscillations cancel each other almost exactly. The delicate balance is easily perturbed, and an attempt to
destroy interference between different durations would also destroy the tunnelling transition one wanted to study. 
\section{Discussion}
Finally, if Alice were to repeat also the experiment of Ref.\cite{Stein}, this is what Bob would 
say about her result.
\e{A fundamental problem, arising each time a Larmor clock is applied to tunnelling, but often overlooked -
the proverbial elephant in the room - has to do with the quantum Uncertainty Principle.
According to the Principle, one can have tunnelling, and not know the time spent in the barrier, or know this duration, 
but have tunnelling destroyed. One faces precisely the same choice in the double slit experiment, where he/she must decide 
between knowing the slit chosen by the particle, or having the interference pattern on the screen, but not both at the same time. 
You have tried to keep tunnelling intact  (your clock perturbs it only slightly), and learn something about the duration spent in the barrier. 
You might expect the UP to make your result {\it always} look flawed in one way or another,  but this is not how the UP works. 
If you consider all possible 
experiments of this type,  some of them will give seemingly reasonable outcomes, whereas other \e{times} would be negative, too short, too long, etc.
This is necessary, and is possible because such \e{times} can be expressed as the combinations of probability amplitudes which, unlike probabilities, have few restrictions on their signs and magnitudes. Though your result of $0.61$ $ms$  does look plausible you cannot recommend using it 
the way you would use a classical time scale just because of this. After all, in a double-slit experiment one cannot cherry pick the points 
on the screen, where the \e{which way?} question can be answered meaningfully, since the Uncertainty Principle applies everywhere in equal measure.
You cannot say that you resolved the controversy regarding how long a tunnelling particle spends in the barrier region, or proved that this duration is non-zero. The controversy, if you wish to call it that, goes to the very heart of the quantum theory, and must be accepted, rather than resolved.}
\section{Methods}
\subsection{Probability amplitude to spend a given duration $\tau$ in the barrier}
Consider a particle with a mean momentum $p_0$, prepared in a wave packet state $G_0$  ($\hbar=1$), 
\begin{eqnarray}\label{a1}
G_0(x) = \int \textcolor{black}{a}(p-p_0)\exp(ipx)dp= \exp(ip_0x)\textcolor{black}{W(x)}, 
\end{eqnarray}
where ${a}(p-p_0)$ discribes the distribution of the particle's momenta, and $W(x)$ is the wave packet's envelope.
At $t=0$ the wave packet lies to the left of a potential barrier $V(x)$ of a width $d$, as shown in the inset in Fig.1. All momenta $p$ in (\ref{a1}) are such that in order to cross the barrier 
the particle has to tunnel. The probability amplitude to detect  the particle at $x$ close to the maximum of the transmitted wave packet, 
after it has been in motion for $\T$ seconds, 
 can be represented as a sum 
over Feynman paths, 
\begin{eqnarray}\label{a2}
A(x\gets G_0) = \int dx' \sum_{paths} \exp(iS[x(t)]G_0(x'), 
\end{eqnarray}
where a path $x(t)$ starts in $x'$ at $t=0$, and ends in $x$  at $t=\T$.
The action functional is given by the usual $S[x(t)]=\int_0^{\T}[{\mu \dot {x}^2}/{2}-V(x)]dt$, with $\mu$ denoting the particle's mass. 
Each path spends a certain amount of time in the barrier region $0\le x\le d$. Thus duration can be computed with the help of a \e{stop-watch} (SW) expression, 
\begin{eqnarray}\label{a3}
\t[x(t)]=\int_0^{\T} \theta_{[0,d]}(x(t)) dt,
\end{eqnarray}
where $ \theta_{[0,d]}=1$ for $0\le x\le d$ and $0$ otherwise, so that only the time intervals spent in the barrier are added to the total.
It is readily seen that $\t[x(t)]$ cannot be negative, nor can exceed the time the particle was in motion, hence
\begin{eqnarray}\label{a3}
0\le \t[x(t)]\le \T. 
\end{eqnarray}  
A simple cosmetic operation turns the path sum (\ref{a2}) into the sum over durations spent in the barrier. Restricting the summation 
to the paths which spend there precisely $\tau$ seconds, yields
\begin{eqnarray}\label{a4}
A(x\gets G_0|\tau) \equiv  \int dx' \sum_{paths} \delta(\t[x(t)]-\tau) \exp(iS[x(t)]G_0(x'), 
\end{eqnarray}
where $\delta(z)$ is the  Dirac delta, and we have
\begin{eqnarray}\label{a5}
A(x\gets G_0) \equiv  \int_0^{\T}A(x\gets G_0|\tau)d\tau.
\end{eqnarray}
This is bad news for one's effort to determine the time {\it actually} spent in the potential - all such durations interfere.
We are back to the Young's interference experiment, except that instead of two paths, each going through one of the slits, we have a continuum of routes, each labelled 
by the value of the $\t[x(t)]$. According to the Uncertainty Principle \cite{FeynL}  the \e{which way?} ( \e{which $\tau$?}) question 
has no answer. The only exception is  the classical limit. Typically, $A(x\gets G_0|\tau)$ is highly oscillatory, but in a classically allowed 
case, e.g., with the barrier removed, the oscillations are slowed down near the classical value $\tau_{cl}=\mu d/p$. If $A(x\gets G_0|\tau)$
has a unique {\it stationary phase point} of this kind, $\tau_{cl}$ will appear as the only time parameter, whenever one  evaluates integrals 
involving $A(x\gets G_0|\tau)$, and classical mechanics will apply as a result. 
\newline
The problem with tunnelling is that no such preferred time emerges for a classically forbidden transition, and all $\tau$s must be 
treated equally (a similar situation is shown in Fig.3 of \cite{DSnat}, although for a different quantity). To make things worse, in tunnelling the amplitude  $A(x\gets G_0)$ is very small ($\sim \exp[-(2\mu V-p_0^2)^{1/2}d]$ for a rectangular barrier), while $A(x\gets G_0|\tau)$ is not. 
Thus, the exponentially small tunnelling amplitude results from a highly accurate cancellation between (not small) oscillations of  $A(x\gets G_0|\tau)$. For this reason, any attempt to modify or neglect any part of the integrand in Eq.(\ref{a4}) would considerably change the result, and destroy the tunnelling. 
\subsection{An uncertainty relation for the duration $\tau$}
Although the Uncertainty Principle hampers one's attempts to ascribe a unique barrier duration to a tunnelling transition, 
there is still one more thing we can do. Writing the $\delta$-function in (\ref{a4}) as
\begin{eqnarray}\label{a6}
\delta(\t[x(t)]-\tau)= (2\pi)^{-1}\int d\lambda \exp\{i\lambda (\tau-\t[x(t)])\}, 
\end{eqnarray}
and inserting it into (\ref{a4}), we note that the new action 
\begin{eqnarray}\label{a7}
S_\lambda[x(t)] \equiv S[x(t)] -\lambda \int_0^{\T}dt  \textcolor{black}{\theta_{[0,d]}(x(t))}
\end{eqnarray}
 corresponds to adding to the barrier $V(x)$ a rectangular potential $\lambda \theta_{[0,d]}(x(t))$, a well or a barrier, depending on the sign of 
 $\lambda$. Equation (\ref{a4}) can now be written in an equivalent form, 
 \begin{eqnarray}\label{a8}
A(x\gets G_0|\tau)=(2\pi)^{-1}\int_{-\infty}^{\infty} d\lambda \exp(i\lambda \tau) \AA(x\gets G_0|\lambda), 
\end{eqnarray} 

where $\AA (x\gets G_0|\lambda)$ is the amplitude to reach, at $t=\T$, the final location $x$ from the initial state $G_0$, 
while moving in a combined potential  $V(x)+\lambda \theta_{[0,d]}(x)$. In other words, to evaluate the amplitude $A(x\gets G_0|\tau)$ 
one needs to know the amplitudes of transmission for all composite potentials. And {\it vice versa}, to know the amplitude for a given potential 
one needs to know the amplitudes for all durations spent therein.
\newline
Note next that even the calculation of the full amplitude distribution of the durations spent in a region $[0,d]$ for a free particle, $V(x)=0$, is already a non-trivial task. 
 It involves evaluation of the transmission amplitudes for all rectangular wells and barriers, and integration in Eq.(\ref{a8}).
However, once $A_0(x\gets G_0|\tau)$ is obtained, the distribution for a rectangular potential $V(x)=V\theta_{[0,d]}(x)$ comes for free, 
 \begin{eqnarray}\label{a9}
A(x\gets G_0|\tau)=\exp(-iV\tau)A_0(x\gets G_0|\tau).  
\end{eqnarray}
As we mentioned above, in the semiclassical limit, the free amplitude distribution $A_0(x\gets G_0|\tau)$ develops a stationary region
around $\tau=\mu d/p_0$. When the  barrier is raised, the factor $\exp(-iV\tau)$ destroys the stationary region, 
$A(x\gets G_0|\tau)$  rapidly oscillates everywhere, and $A(x\gets G_0)$ becomes small for a tunnelling particle. 
\newline
Equation (\ref{a8}) is a kind of uncertainty relation between the duration $\tau$ and the potential in the region of interest. 
It implies that a device employed to measure the $\tau$ must introduce some uncertainty into the potential,  
the greater the uncertainty, the more accurate the measurement. Which brings us to the Larmor clock. 
\subsection{The Larmor clock}
The clock consists of a magnetic moment, proportional to an angular momentum (spin) of a size $j$, coupled to a magnetic field along the $z$-axis 
via $\h_{int}=\om \hat j_z$, where $\om$ is the Larmor frequency. By the time $t$, an initial state 
 \begin{eqnarray}\label{a10}
|\gamma\ra=\sum_{m=-j}^j\gamma_m|m\ra, \q \hat j_z|m\ra =m|m\ra
\end{eqnarray}
becomes rotated by an angle $\om t$ around the $z$-axis, 
 \begin{eqnarray}\label{a11}
|\gamma(t)\ra=\exp(-i\om t\hat j_z)|\gamma(0)\ra = \sum_{m=-j}^j\gamma_m\exp(-i m\om t)|m\ra. 
\end{eqnarray}
Suppose the spin travels with a classical particle moving along a trajectory $x(t)$, and the field exists only in the region $0\le x \le d$.
Then the spin,  precessing only when the particle is in the field, $0\le x(t) \le d$, ends up rotated by $\om \t[x(t)]$ by $t=\T$. 
 Quantally, for a particle in the inset of Fig.1, the final (unnormalised) spin's state can be found simply by adding up its states, rotated by $\om \tau$, 
 each multiplied by the probability amplitude of spending in the field a net duration $\tau$. The result is 
 \begin{eqnarray}\label{a12}
|\gamma(\T)\ra=\int_0^{\T}d\tau A(x\gets G_0|\tau)\exp(-i\om \tau \hat j_z)|\gamma(0)\ra. 
\end{eqnarray}
In general,  the r.h.s. of (\ref{a12}) cannot be rewritten as a single rotation around the $z$-axis by an angle $\om \tau'$, $|\gamma(\T)\ra \ne \exp(-i\om \tau' \hat j_z)|\gamma(0)\ra$, 
and no unique time $\tau'$ can be associated with a quantum transition in this way.
\newline
With the help of Eq.(\ref{a8}),  one obtains an equivalent form of Eq.(\ref{a12}), 
 \begin{eqnarray}\label{a13}
|\gamma(\T)\ra= \sum_{m=-j}^j\AA(x\gets G_0|m\om)\gamma_m|m\ra. 
\end{eqnarray}
This shows that each spin component traverses the barrier as if the potential there were $V(x)+m\om$, so the potential, experienced by the 
particle as a whole, remains uncertain within  the range from $-j\om$ to $j\om$. As was already noted, a viable clock has to introduce this uncertainty, and
we may ask what can be learnt about the duration spent in the barrier by applying the Larmor clock.
\newline
An experiment could consist in detecting, at $t=\T$,  the particle in $x$ and its spin in a state $|\beta\ra=\sum_{m=-j}^j\beta_m|m\ra$.  From (\ref{a12}) the corresponding 
probability is 
  \begin{eqnarray}\label{a14}
P(\beta, x \gets \gamma,G_0) = |\int_0^{\T} d\tau \Gamma(\tau|\om,\beta,\gamma)A(x\gets G_0|\tau)|^2, 
\end{eqnarray}
where 
  \begin{eqnarray}\label{a15}
\Gamma(\tau|\om,\beta,\gamma)\equiv \la \beta |\exp(-i\om \tau \hat j_z)|\gamma\ra=\sum_{m=-j}^j\beta^*_m \gamma_m\exp(-i m\om \tau).
\end{eqnarray}
Thus, by measuring the probability (\ref{a14}), one can determine the absolute value of the integral in Eq.(\ref{a14}), which involves the amplitude distribution of the durations spent by the particle inside the barrier
in the absence of the clock. Note that little is left of the original tunnelling transition, where the transmission  amplitude $A(x\gets G_0)$ 
is typically small. As already mentioned at the end of subsection A, the presence of an additional factor such as $\Gamma(\tau|\om,\beta,\gamma)$ is likely to alter destructive interference which defines tunnelling. As a result,  $\int_0^{\T} d\tau \Gamma(\tau|\om,\beta,\gamma)A(x\gets G_0|\tau)$ could differ from the original tunnelling amplitude in Eq.(\ref{a5}) by orders of magnitude.
\subsection{A non-perturbing (weak) Larmor clock }
One can try to return to tunnelling by sending $\om \to 0$, and learn something about the tunnelling time from the particle's response to the clock.
(This already bodes ill for one's task, since the uncertainty introduced in the potential will also tend to zero, which, according to subsection B, 
should lead to a large uncertainty in $\tau$). Nevertheless, we obtain
  \begin{eqnarray}\label{a16}
\Gamma(\tau|\om,\beta,\gamma) \approx \la \beta|\gamma\ra - i\om \tau \la \beta|\hat j_z|\gamma\ra,
\end{eqnarray}
so that the relative change in the probability (\ref{a14}) with and without the magnetic field is
  \begin{eqnarray}\label{a17}
\frac{P(\beta, x \gets \gamma,G_0)_{\om}-P(\beta, x \gets \gamma,G_0)_{{\om}=0}}
{P(\beta, x \gets \gamma,G_0)_{{\om}=0}}
\approx 
2 \R [Z(\beta, \gamma)]\I[\overline \tau]+ \I[Z(\beta, \gamma)]\R[\overline \tau],
\end{eqnarray}
where $Z(\beta, \gamma)\equiv {\la \beta|\hat j_z|\gamma\ra/\la \beta|\gamma\ra}$ and 
  \begin{eqnarray}\label{a18}
\overline \tau \equiv \frac {\int_0^{\T} \tau A(x\gets G_0|\tau) d\tau}{\int_0^{\T} A(x\gets G_0|\tau) d\tau}
=\frac {\int_0^{\T} \tau A(x\gets G_0|\tau) d\tau}{A(x\gets G_0)}
\equiv \int_0^{\T} \tau \alpha(\tau) d\tau
\end{eqnarray}
is the complex time of Sokolovski and Baskin \cite{SB}. 
The quantity in the l.h.s. of Eq.(\ref{a18}) can be measured, and by choosing a different $|\beta\ra$ one can, in principle, determine the values
of $\R[\overline \tau]$, $\I[\overline \tau]$, or indeed of their various combinations. Moreover, for $ \la \beta|\gamma\ra=0$, one has
 \begin{eqnarray}\label{a19}
P(\beta, x \gets \gamma,G_0)_{\om}\sim \om^2|\overline \tau|^2,
\end{eqnarray}
so the modulus of $\overline \tau$ can also be determined directly. 
\newline
Now there are many real valued time parameters related to the complex time (\ref{a18}), yet none of them is a suitable candidate 
for a physical time interval representing the net duration spent in the barrier. The easiest way to demonstrate it is to note that for an improbable transition, $A(x\gets G_0|\tau) \to 0$,  
the denominator of (\ref{a18}) can be very small. At the same time, the numerator does not have to be small, since multiplication of $A(x\gets G_0|\tau)$ by $\tau$ can destroy the cancellation, characteristic of tunnelling. Thus, $|\overline \tau|$ may, in principle, exceed the total duration of motion, 
$|\overline \tau| >> \T$. This makes little sense, especially if one recalls that each and every Feynman path in Eq.(\ref{a2})  spends in the barrier no more than $\T$. 
\subsection{The Baz' clock}
Finally we briefly discuss a particular type of a weak Larmor clock, employing a spin-$1/2$ in a weak magnetic filed. It was introduced by A.I. Baz' more than fifty years ago \cite{Baz}, and recently implemented by Ramos {\it et al} in \cite{Stein}. Now $\hat j_z = \sigma_z/2$ ($\sigma_z$ is the Pauli matrix), and the spin's initial direction is along the $x$-axis, whose azimuthal and polar angles  are $\phi=0$ and $\theta=\pi/2$ respectively. 
According to (\ref{a12})  the final (unnormalised) state of the spin is given by 
 \begin{eqnarray}\label{a21}
 |\gamma(\T)\ra=2^{-1/2}
 \begin{pmatrix}
\int_0^{\T}d\tau A(x\gets G_0|\tau)\exp(-i\om \tau/2)\\
\int_0^{\T}d\tau A(x\gets G_0|\tau)\exp(+i\om \tau/2)
 \end{pmatrix}\n
 \approx 2^{-1/2}A(x\gets G_0|\tau)
 \begin{pmatrix}
1-i\om\R [\overline \tau]/2 +\om\I [\overline \tau]/2  \\
1+i\om\R [\overline \tau]/2 -\om\I [\overline \tau]/2)
 \end{pmatrix}.
\end{eqnarray}
As it was discussed in subsection C, this cannot in general correspond to a rotation around the $z$-axis. 
On the other hand, in {any} state, a spin-$1/2$ must point along some direction on the Bloch sphere. 
Thus, we expect the state (\ref{a21}) to be rotated not only in the $xy$-, but also in the $xz$-plane.
The state of a spin, polarised along a direction 
making angles $\delta \phi$ and $\pi/2-\delta \theta$
with the $x$- and the $z$-axis, respectively, can be written as
 \begin{eqnarray}\label{a22}
 \begin{pmatrix}
\exp(-i\delta \phi/2) \cos\left (\frac{\pi}{4}-\frac{\delta \theta}{2}\right)\\
\exp(+i\delta \phi/2) \sin\left (\frac{\pi}{4}-\frac{\delta \theta}{2}\right)
 \end{pmatrix}
 \approx  2^{-1/2}
 \begin{pmatrix}
1-i\delta \phi/2+\delta \theta/2\\
1+i\delta \phi/2-\delta \theta/2) 
 \end{pmatrix}.
\end{eqnarray}
Comparing (\ref{a21}) with (\ref{a22}) we find that the spin has rotated by the (small) angles 
 \begin{eqnarray}\label{a23}
\delta \phi= \om \R[\overline \tau], \q \text {(in the $xy$-plane)} \q \text{and} \q \delta \theta= \om \I[\overline \tau] \q \text {(in the $xz$-plane)}.
\end{eqnarray}
We recall further that a spin travelling with a {\it classical} particle along a trajectory $x_{class}(t)$ 
would rotate {\it only} in the $xy$-plane by an angle $\om \tau_{class}= \t[x_{class}(t)]$.  
Thus, the first of Eqs.(\ref{a23}) looks like the classical result, with $ \tau_{class}$ replaced by $\R[\overline \tau]$.
The second of  Eqs.(\ref{a23}) has no classical analogue, and should serve as a warning that a straightforward 
extension of the classical duration to the quantum case may not be possible. (One already knows this from the Uncertainty 
Principle.)
\newline
The appearance of not one, but two rotation angles was first noted by B\"uttiker in \cite{Butt}, albeit in a slightly different context.
[Ref. \cite{Butt} considered transmission of a particle with a known momentum $p_0$ which, in our language, corresponds to replacing 
$A(x\gets G_0|\tau)$ with $A(p_0\gets G_0|\tau)\equiv \int \exp(-ip_0x)A(x\gets G_0|\tau)dx$ in all formulae, and making $G_0$ nearly monochromatic.]
In \cite{Butt} B\"uttiker defined two \e{times},  $\tau_y \equiv \delta \phi/\om$ and $\tau_z \equiv \delta \theta/\om$, which correspond to our
$R[\overline \tau]$  and $\I[\overline \tau]$, respectively. Ramos {\it et al} measured both the real and the imaginary parts of $\overline \tau$, which can be seen in Fig.3 of \cite{Stein}. 
The authors of \cite{Stein} found both parameters positive and concluded that their results were \e{inconsistent with claims that tunnelling takes 'zero time'}. To abide by this conclusion one needs to take for granted that the "time tunnelling takes" exists as a meaningful concept, 
but this is not the case. 
\newline
The confusion can be traced back to B\"uttiker \cite{Butt}. When faced with two times parameters instead of one, 
he opted for a non-negative combination of the two, $\tau_x \equiv \sqrt{\tau_y^2+\tau_z^2}$. This equals
the modulus of the \e{complex time} in \textcolor{black} {Eq.(\ref{a18})}, $\tau_x=|\overline \tau|$.
At least one point made in \cite{Butt} requires a comment, if not a correction. In $\tau_x$ B\"uttiker believed to have found  (we read in the Abstract of \cite{Butt})
\e{the time interval during which a particle interacts with the barrier if it is finally transmitted.} However, 
neither $\R[\overline \tau]$ nor $\I[\overline \tau]$, nor any combination of the two can be interpreted as a physical time interval.
A weighted sum of quantum mechanical amplitudes, $\overline \tau$, may not give a meaningful answer to the question \e{how much time 
does a tunnelling particle spend within the barrier region?} for the same reason the Uncertainty Principle \cite{FeynL} forbids 
identifying the particle's path in Young's double-slit experiment.

 \begin{center}
\textbf{Acknowledgements}
\end{center}
Financial support of
MCIU, through the grant
PGC2018-101355-B-100(MCIU/AEI/FEDER,UE) (DS), and of the Basque Government through Grant No IT986-16 (DS) is gratefully acknowledged. EA acknowledges the financial support of the Ministerio de Econom\'ia y Competitividad (MINECO) of the Spanish Government through BCAM Severo Ochoa accreditation SEV-2017-0718 and PID2019-104927GB-C22 grant. This work was also supported by the BERC 2018e2021 Program and ELKARTEK Programme (KK-2020/00049 and KK-2020/00008) funded by the Basque Government. 
\begin{center} 
\textbf{Author contributions}
D.S. and E.A. both wrote the paper, and reviewed it.
\end{center}
\begin{center} 

\textbf{Competing financial interests}
The authors declare no competing financial interests.
\end{center}
\begin{center} 

\textbf{Corresponding authors}
Correspondence to Dmitri Sokolovski, {dgsokol15@gmail.com}.
\end{center}
\end{document}